\documentclass[preprint]{aastex}
\usepackage{times}
\usepackage{natbib}
\usepackage{graphicx}

\newcommand{\eb}[1]{\begin{equation}\label{eq:#1}}
\newcommand{\en}{\end{equation}}
\newcommand{\eqa}[1]{\begin{eqnarray}\label{eq:#1}}
\newcommand{\eqn}{\end{eqnarray}}

\newcommand{\oh}[1]{\omit\hidewidth #1\hidewidth}

\newcommand{\dd} [2]{\frac{{\rm d}{#1}}{{\rm d}{#2}}}
\newcommand{\up}{\uparrow}
\newcommand{\dn}{\downarrow}

\shortauthors{Trampedach and Stein}

\begin{document}

\title{The Mass Mixing Length in Convective Stellar Envelopes}
\author{Regner Trampedach}
\affil{JILA, University of Colorado, Boulder, CO 80309-0440, USA}
\email{trampeda@lcd.colorado.edu}
\author{Robert F. Stein}
\affil{Department of Physics \& Astronomy,
 Michigan State University, East Lansing, MI 48824, USA}
\email{stein@pa.msu.edu}

\begin{abstract}
The scale length over which convection mixes mass in a star can be calculated
as the inverse of the vertical derivative of the unidirectional
(up or down) mass flux.  This is related to the mixing length in
the mixing length theory of stellar convection.  We give the ratio of mass
mixing length to pressure scale height for a grid of 3D surface convection
simulations, covering from 4300\,K to 6900\,K on the main-sequence, and
up to giants at $\log g = 2.2$, all for solar composition.
These simulations also confirm what is already known from solar simulations, 
that convection doesn't proceed by discrete convective elements, but rather
as a continuous, slow, smooth, warm upflow and turbulent, entropy deficient,
fast down drafts. This convective topology also results in mixing on a scale
as that of the classic mixing length formulation, and is simply a consequence
of mass conservation on flows in a stratified atmosphere.
\end{abstract}


\section{Introduction}

For stellar structure and evolution calculations \citep[e.g.,][]{Iben67},
models for the relevant energy-transport mechanisms are needed. In the
interior, radiative or conductive transfer of energy is trivial, although the
atomic physics going into the opacities and conductivities are not.
A satisfactory description of convection in a manner amenable to stellar
structure calculations, however, has proven elusive.  Stellar modelers 
usually use the mixing length theory (MLT)
\citep{Vitense58,gough:state-of-MLT}, which despite its simplicity and many shortcomings,
has been highly successful in matching a variety of stellar observations.
In this model, convection is thought of as a
collection of fluid parcels that travel a distance
equal to the mixing length and then thoroughly mix with the surrounding
ambient medium, sharing their energy. The MLT picture is completely symmetric,
so the upward flow of buoyant warm elements is accompanied by an equal number of
cooler and over-dense elements moving inward.
In these calculations the
mixing length is a free parameter and is specified as a multiple
of the pressure scale height $H_P=P/\rho g$, where $P$ is the pressure,
$\rho$ is the density and $g$ is the gravitational acceleration.
In general, there are also free parameters specifying the geometry
of the fluid parcel -- the ratio of its horizontal to vertical size
and another specifying the lateral radiative transfer between
convective elements and the ambient medium.
The simple mixing length model is used due to a lack of better alternatives
and because it is quick to
compute.  Its primary function is to determine the entropy of the
deep, efficient and nearly adiabatic part
of the convection zone in the star.  If the magnitude
of the mixing length and the other parameters could be specified a priori, simple mixing
length theory would be capable of fulfilling this function.  However,
there is little guidance from theory as to their values.

In addition, the mixing length model suffers
from several other problems: it does not allow for overshooting
motions that mix material into the surrounding stable layers, 
it does not allow for turbulence,
it does not allow for asymmetry between upflows and downflows, especially
the large temperature difference near the surface and its effect on opacities
and radiative
transfer.
These are all phenomena that,
according to realistic
3D simulations, contribute to the atmospheric entropy jump and hence, the
adiabat of the deep, efficient convection.
Several attempts have been made to improve the mixing length model.
Some have taken the mixing length to be the distance to the nearest
boundary \citep{Stothers97}, some have introduced a two-stream model
with separate properties of the upflows and downflows \citep{Nordlund76},
some have introduced the idea of a spectrum of sizes of the convective
cells \citep{Canuto91}.

Several attempts have been made to determine the appropriate value
of the ratio of mixing length to pressure scale height by matching
theoretical and observed evolutionary tracks in the Hertzsprung-Russell
diagram \citep{Chieffi95,Stassun04}, by matching observed and model
properties of binary systems \citep{Fernandes98,Fernandes02,Montalban04},
and red giant temperatures \citep{Ferraro06}.  There have also been
a few attempts to determine the mixing length parameter by comparing
one-dimensional mixing length stellar model calculations with two-
and three-dimensional numerical simulations of stellar convection
by either matching entropy in the adiabatic portion of the convection
zone \citep{Ludwig99,Freytag99} or by matching temperature
and density at a given pressure in the interior
\citep{Trampedach99,Trampedach11}.

Mixing length theory works reasonably well for determining the
thermal structure of convection zones because real mixing of mass
occurs in stratified convection.  Stars are stratified.  Their
density decreases from their center to their surface in order to
produce a pressure gradient to balance the inward pull of gravity.
In order to conserve mass in such a stratified medium, most of the
ascending material must turnover and head back down within about a
density scale height \citep{bob:conv-topology}. Thus, mixing really does occur
on the order of a density scale height.  

It is possible to easily and unambiguously calculate the mass mixing
length from computational fluid dynamic simulations of stellar
convection.  Such simulations, when they employ realistic equations
of state  to relate the pressure to the density and internal energy
or temperature and employ sufficiently accurate treatment of
radiative transfer to determine the heating and cooling, have as
free parameters only the resolution of the computational grid used
and the magnitude of the diffusion used to stabilize the calculations.
These parameters, however, are not free to be adjusted to match observations.
The diffusion is minimized against the requirement of a numerically stable
simulation, and the resolution is chosen as a trade-off between resolving
features in the photosphere and computational cost. We would not lower the
resolution or increase the diffusion to give a better match
to some observations.

In this paper, we present results of such mass mixing calculation for a homogeneous
sample of convection simulations, covering a large part of the
Hertzsprung-Russell diagram for which stars have
convective envelopes. The simulations form an irregular grid, covering
4\,500--6\,900\,K on the main sequence and going up to giants with
$\log g=2.2$ and $T_{\rm eff}=$4\,300-5\,000\,K, 
all with solar composition.
The simulations employ the Mihalas-Hummer-D{\"a}ppen equation of state
\citep{mhd1}, line-opacity from \citet{kur:line-data} and continuum opacities
largely from the Marcs stellar atmosphere package \citep{gus:modgrid}, but
with many improvements. The simulations are described in more detail in
\citet{trampedach:T-tau,Trampedach11}.

\section{Mass Mixing Length}

\begin{figure}[!htb]
\centerline{\includegraphics[width=0.7\textwidth]{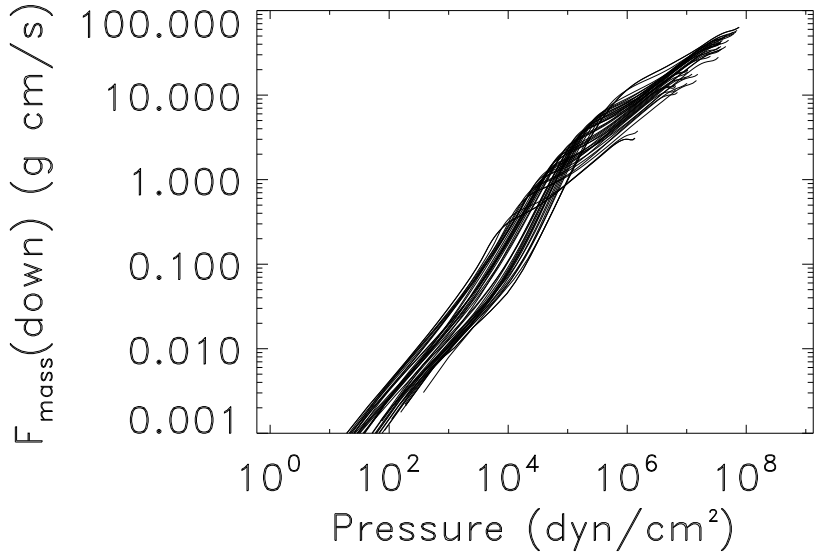}}
\caption{Uni-directional downward mass flux as a function of the mean
pressure for the stars in our sample.  The break in the mass flux near
pressures of $10^5$\,dyn/cm$^2$ occurs in the photosphere.  The effect of the
bottom boundary can be seen as a small downturn near the high pressure
end of the traces.
} 
\label{fig:fmdown} 
\end{figure}

The mass mixing length, the distance over which, because of mass 
conservation, most of the ascending fluid turns over and starts
to descend, is the inverse of the logarithmic
derivative of the unidirectional mass flux \citep{Stein09},
\eqa{eq:massmix}
\ell_m(r) & = & \left|{\rm d} \ln F_{\up m}(r) / {\rm d}r \right|^{-1}\nonumber\\
     & = & \left|{\rm d} \ln F_{\dn m}(r) / {\rm d}r \right|^{-1}
\ .
\eqn
This quantity is readily determined from three-dimensional,
stellar convection simulations.  
The equality of the mass flux in the up- and downflows, is just another way
of stating mass conservation at each depth.

In the rest of this paper we will look at the upflow, which mass flux is
\eb{massfluxcomponents}
    F_{\up m}(r) = \langle\rho(r) V(r)\rangle_\up A_\up(r)
        \approx \langle\rho(r)\rangle_\up \langle V(r)\rangle_\up A_\up(r) \ ,
\en
The brackets, $\langle f(r)\rangle_\up$, denote horizontal averages over the upflows of a quantity $f$:
$\langle f(r)\rangle_\up=\int_{A_\up(r)}f(\vec{s},r) {\rm d^2} \vec{s}/A_\up(r)$,
with $A_\up(r)$ being the area with upflows and $\vec{s}$ is the coordinates in
the horizontal directions. The approximation of the right-hand-side
of Eq.~(\ref{eq:massfluxcomponents}) ignores
correlations between density and velocity, but we
only use this approximation to gain some insights into the
behavior of $\ell_m(r)$, by re-writing it as
\eb{Hterms}
    \ell_m(r) \approx \left|\dd{\ln\langle\rho\rangle_\up}{r}
                     + \dd{\ln\langle V\rangle_\up}{r}
                     + \dd{\ln A_\up}{r}\right|^{-1}\ .
\en
Below the surface region, the area of the upflows is nearly
constant at about 2/3 of the total area, so its gradient has little
effect on $\ell_m$.
From Eq.~(\ref{eq:Hterms}) we see that, ignoring gradients in velocity and
filling factor, the mass mixing length is simply the density scale-height,
$H_\rho = {\rm d}r/{\rm d}\ln\rho$. Another way to look at it, is to
consider an upflow along the decreasing density of an average stratification
with gradient $\dd{\ln\rho}{r}$. If the upflow had retained its density, it
would be overdense by $\Delta\rho/\rho =-\dd{\ln\rho}{r}\Delta r$, compared to
its surroundings, after having traveled the distance $\Delta r$ (remember that
the gradient is negative).  This is obviously unstable, and instead the
fraction $\Delta\rho/\rho$ overturns into the downflows, in order to keep the
hydrostatic balance. That fraction is 1 for $\Delta r=
1/\dd{\ln\rho}{r}=H_\rho$ in a linear stratification and 1/e in an exponential
stratification. Locally
the upflow will therefore be ``eroded'' with an $e$-folding
scale of $H_\rho$. This ``erosion'' of the upflows does not mean that they
loose coherency and disappear over that length-scale --- on the contrary. We
observe coherent upflows over the entire 5.3--7.3 pressure scale heights
covered by the convection zones of our simulations (a similar number of
pressure scale heights are included above the convection zones of the
simulations). We likewise see downdrafts that reach all the way to the bottom
of our simulations (in all of our simulations), with weaker ones either
stopping or merging into fewer strong downdrafts, before they reach the
bottom. The mass mixing length simply describes how only a very small fraction
of the mass in the upflows make it from the interior and all the way up to the
photosphere, simply due to mass conservation in a stratified atmosphere.

\begin{figure}[htb]
\centerline{
\includegraphics[width=0.5\textwidth]{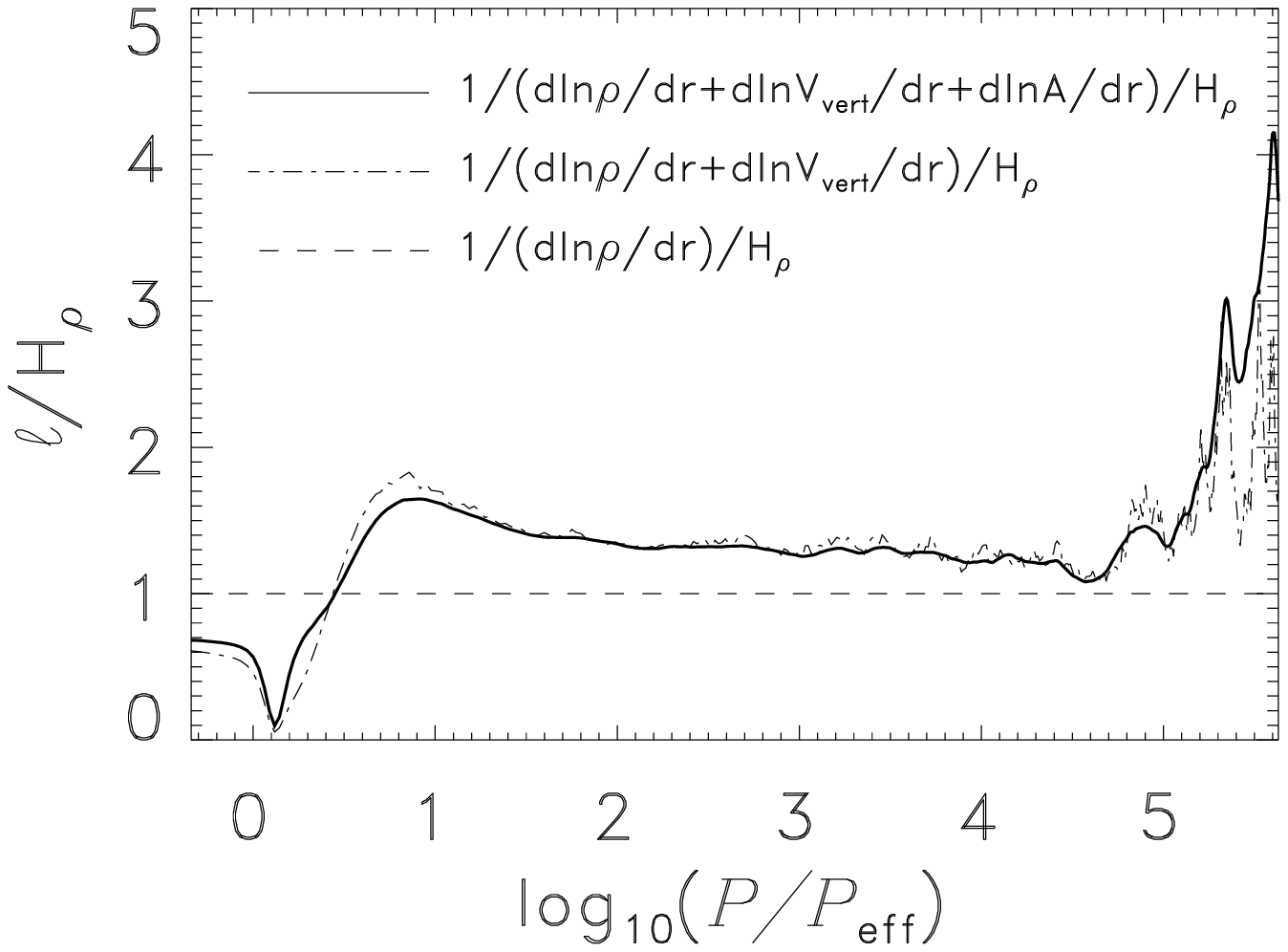}\\
\includegraphics[width=0.5\textwidth]{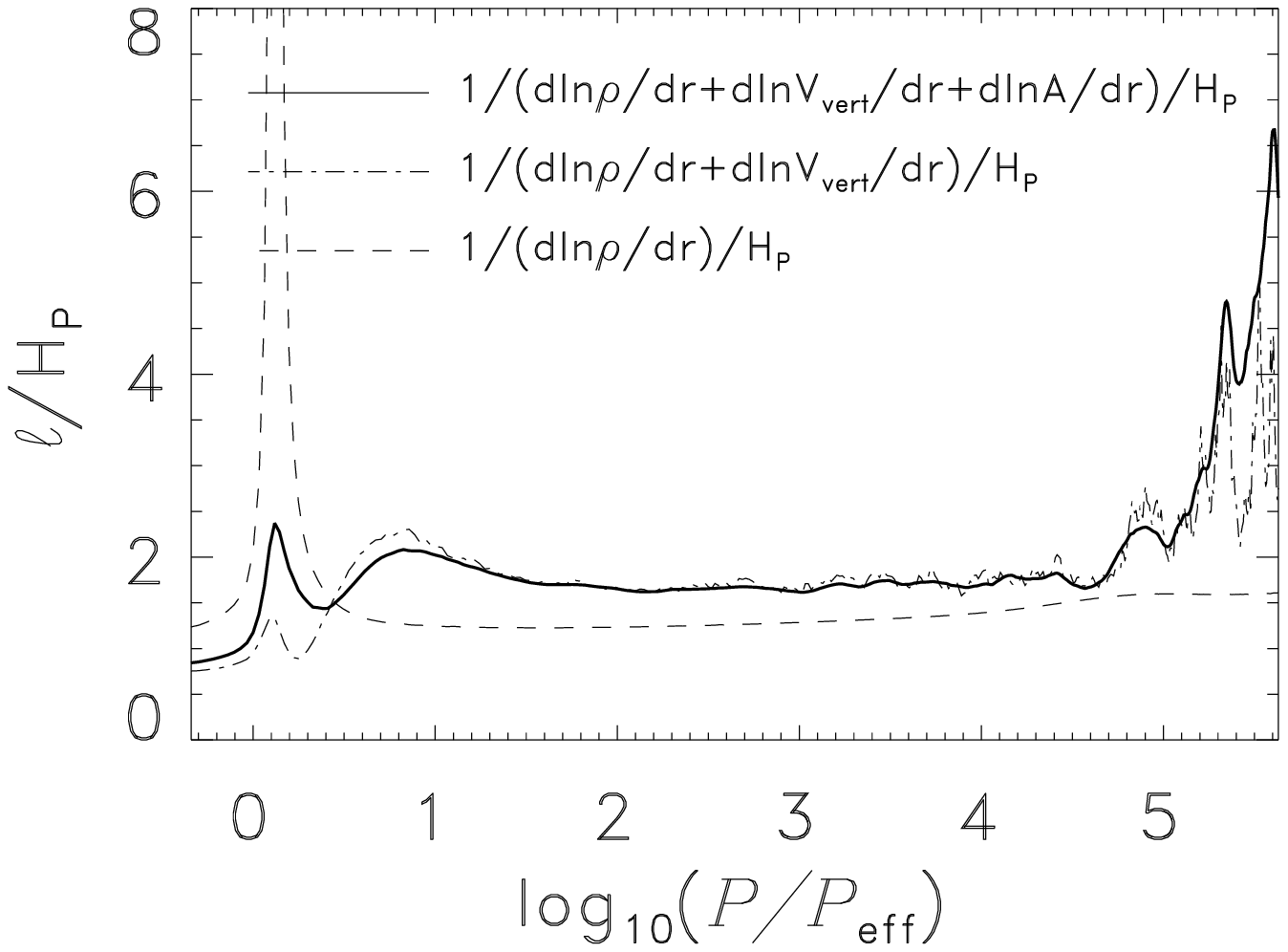}}
\caption{The three terms of Eq.~(\ref{eq:Hterms}) constituting the mass mixing
lengths, in terms of the density scale height, $H_\rho$ (left), and the
pressure scale height, $H_P$ (right). The full expression is shown with solid
line. The effect of ignoring the gradient in filling factor is shown with
dot-dashed, and just including the density gradient results in the dashed line.
The example is shown for the deep solar simulation, also presented in
Fig.~\ref{fig:sun}} 
\label{fig:Hterms} 
\end{figure}

Realistically the velocity decreases inward from the top of the convection zone,
so its derivative is of
opposite sign to that of the density.  This decreases the variation
of the unidirectional mass flux and increases its scale length compared to the
density scale height. In a deep solar simulation, we find a logarithmic
derivative of the upward mass flux of $\simeq 4/(5H_\rho)$
(Figure~\ref{fig:sun}), and the vertical
velocity must therefore vary approximately as
\eb{vz}
    V_{\rm vert} \propto \rho^{-1/5} \ .
\en
The various terms of Eq.~(\ref{eq:Hterms}) are shown for the deep solar
simulation in Figure~\ref{fig:Hterms}. We clearly see that the density gradient
is the main term and that the velocity gradient produces the deviation
from $\ell_m=H_\rho$. The gradient of the filling factor is
only significant in the surface layers. The difference between the approximation
of the right-hand-side of Eq.~(\ref{eq:Hterms}) and the full expression, can
be seen by comparison with Figure~\ref{fig:sun}, which is based on the latter.

Near the surface, the mass mixing length is not a constant multiple of
either the pressure or density scale heights (Figs. \ref{fig:Hterms}
and \ref{fig:sun}). This is due to the abrupt changes in all three
mass-flux ingredients in the surface layers.  The density gradient
becomes very small (for all but the coolest dwarfs the density in the
upflows has a sub-photospheric inversion) resulting in a spike in the
density's contribution to the mixing length (See right hand panel
of Fig.~\ref{fig:Hterms}).  The small density gradient is however
accompanied by a large velocity gradient due to buoyancy braking of
the upflows where the density gradient becomes small or inverted, which
reduces the density induced peak in the mixing length.  In the interior
the density and velocity gradients have opposite signs and a nearly
constant ratio.  However, as the surface is approached the buoyancy
driving increases, which steepens the velocity gradient and gives rise to
the wide sup-photospheric bump in the mixing length.  The photospheric
side of this bump is considerably smoothed by the contribution from
the filling factor.  Division by the pressure scale height to get the
$\ell/H_P$ presented in Figs.~\ref{fig:set1}-\ref{fig:set2} decreases
the difference between the atmospheric and the interior values.

To determine the mass mixing length, $\ell_m$, for a simulation, we first calculate
the horizontal and time averages of the vertical upward (and downward)
momentum (which is the unidirectional mass flux).  The unidirectional
mass flux increases with increasing depth as the density increases
and the velocity decreases, but more slowly than $\rho^{-1/2}$.  We
then calculate the inverse of the logarithmic derivative of this
total up- or downward unidirectional mass flux.  
We have computed the mass mixing length for the set of realistic, compressible,
convection simulations by \citet{Trampedach11}.  Figure~\ref{fig:fmdown}
shows the unidirectional (downward) mass flux for all the stars in 
our sample.  Next we determine the multiple of the
pressure and density scale heights that best match the inverse
logarithmic derivative of the unidirectional mass flux between its
local maximum at the steep temperature gradient near the surface
and where the bottom boundary conditions begin to affect it,
due to boundary effects on the velocity, seen as the upturn at the
bottom in Figs.~\ref{fig:sun}--\ref{fig:set2}.  This increase in
the mass mixing length near the bottom boundary is an artifact of
the bottom boundary condition that the upflows are forced towards
vertical at the boundary.

\section{Results}

\begin{figure}[htb]
\centerline{\includegraphics[width=0.33\textwidth]{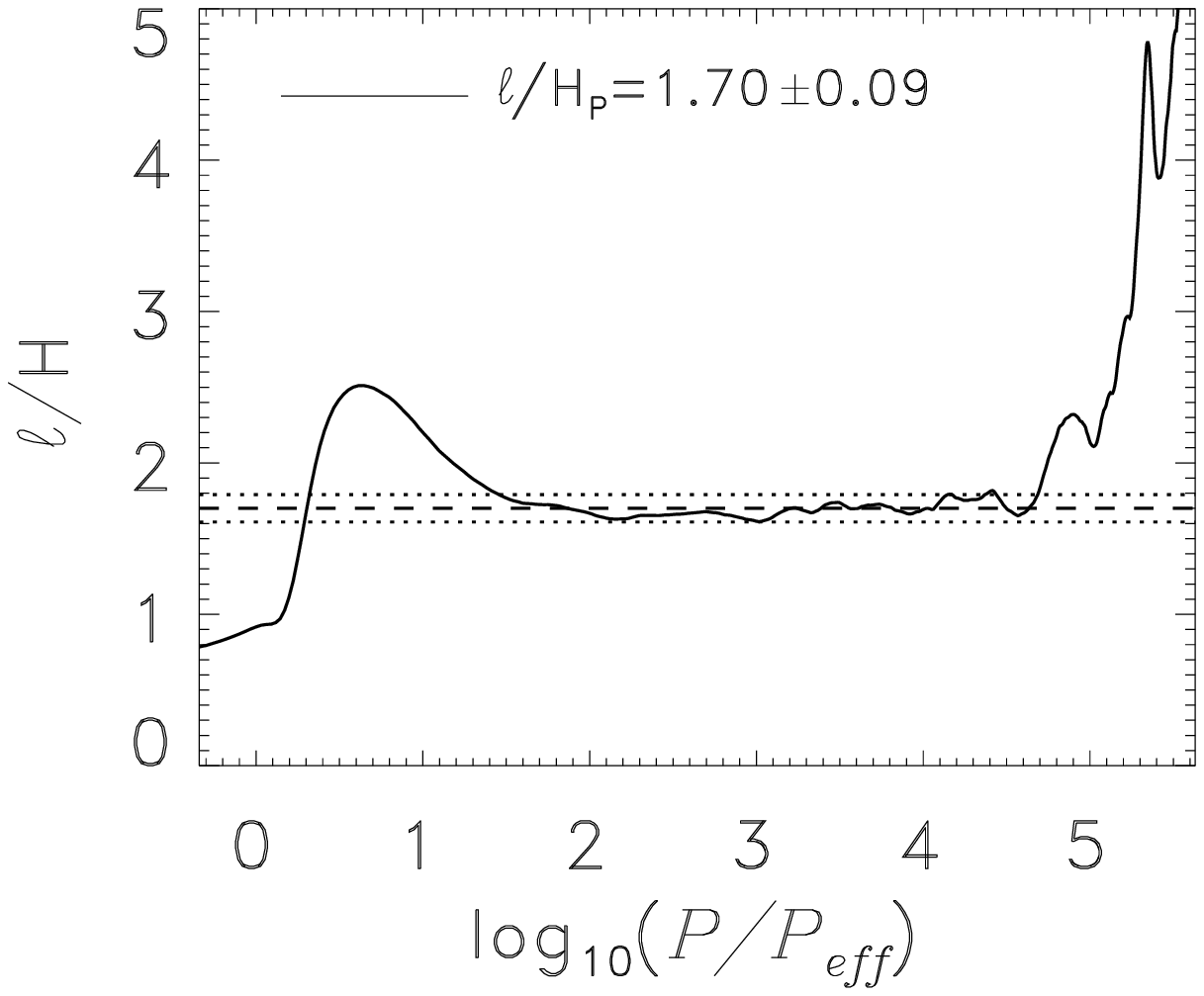}\\
\includegraphics[width=0.33\textwidth]{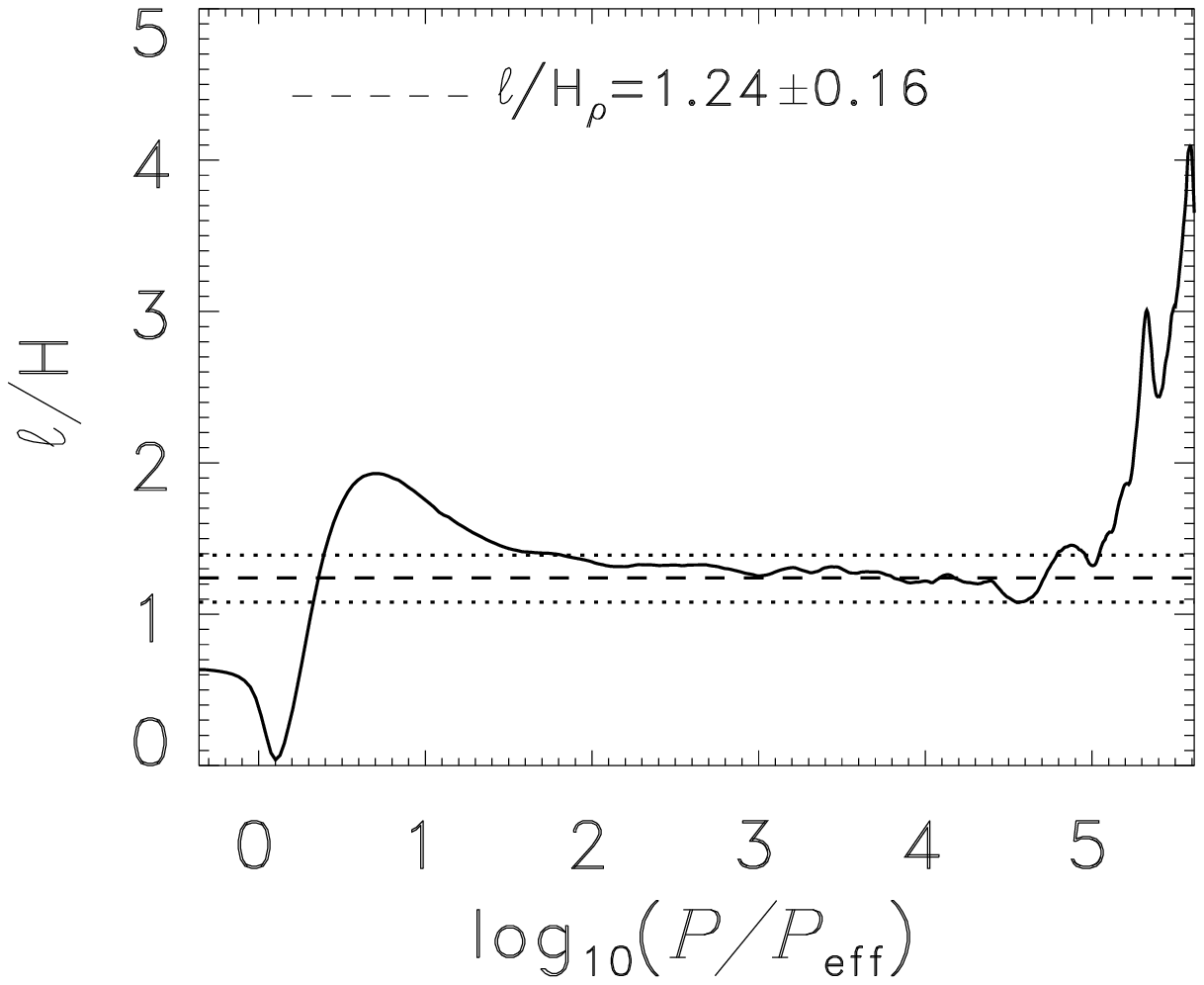}\\
\includegraphics[width=0.33\textwidth]{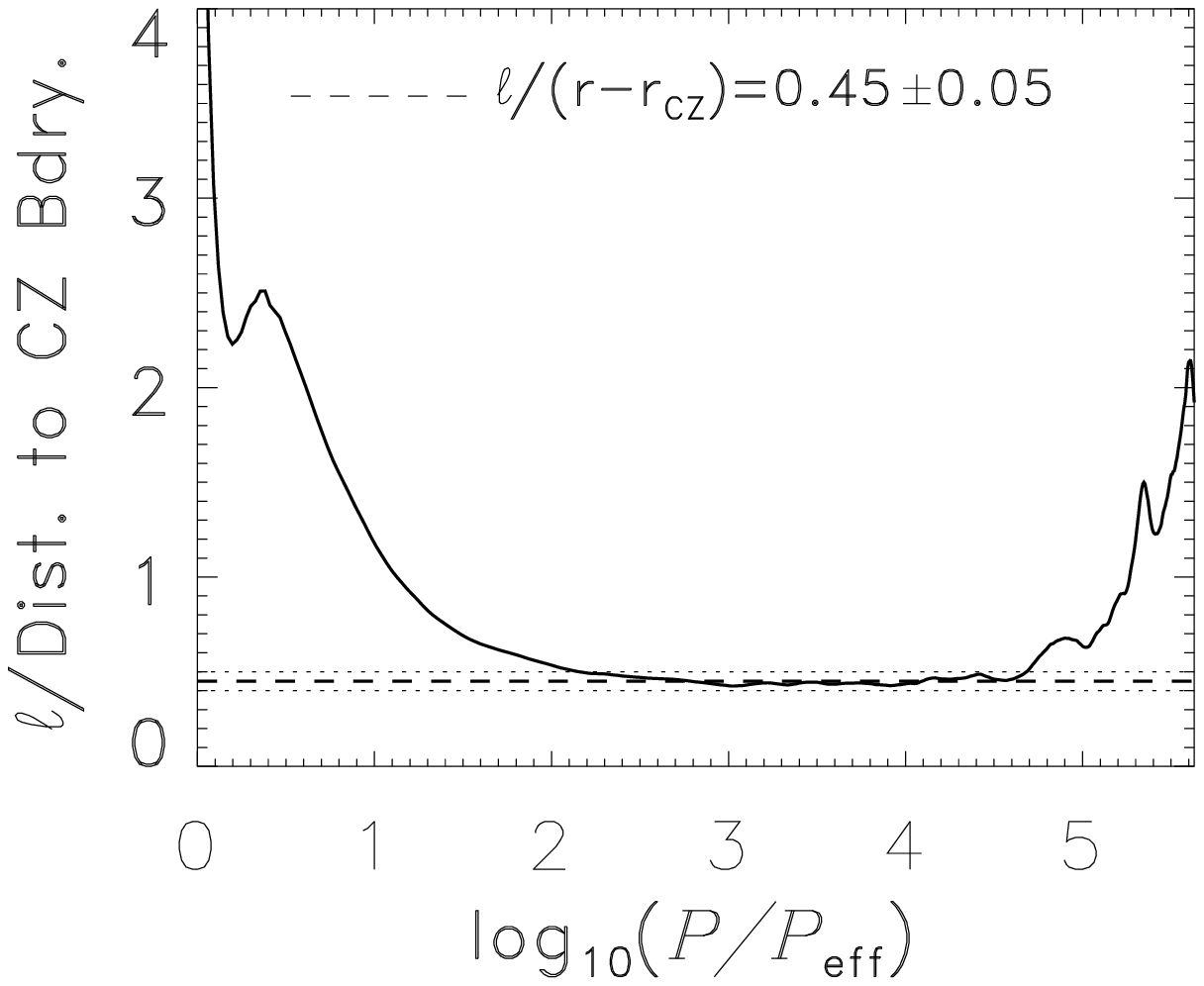}} 
\caption{Sun's
mass mixing length (solid) in units of the pressure scale height (left),
density scale height (middle) and distance to the top of the convection zone
(right) as a function of the mean pressure in units of the surface pressure.
Dashed line is the average and dotted lines are the extremes away from the
boundaries.} 
\label{fig:sun} 
\end{figure}
 
For the Sun we have deep (20\,Mm) convection calculations \citep{Stein09} and
the determination of the mass mixing length is straightforward, as shown in
Figure~\ref{fig:sun}. From this we see that the ratio of the mass mixing
length to the pressure scale height, $\alpha_P$, is nearly constant with depth
in the interior of the convection zone until the bottom boundary is approached.
The ratio to the density scale height, $\alpha_\rho$, is closer to one, but
varies slightly and linearly with depth.  
The mixing length is greater than the density scale height because buoyancy 
is continually accelerating the flow resulting in a significant contribution
from the second term in Eq.~(\ref{eq:Hterms}).
The mass mixing length in units of the distance to the top of the convection
zone, $\alpha_{r-r_{\rm CZ}}$, is also nearly constant with depth in the
interior, but has a value of only 0.45 rather than one.
In all three cases the constant
proportionality is destroyed at the same point near the bottom boundary due to boundary effects
(See Fig.~\ref{fig:fmdown}).
At the top of the
convection zone, in the highly superadiabatic layer, the mass mixing length
becomes large compared to all three measures.
Hence, larger values of the mixing length parameters, $\alpha_X$, are needed
near the surface than in the interior, with $\alpha_{r-r_{\rm CZ}}$ becoming
largest of all, a factor of 5, compared to only a factor of 1.5 for the two
other methods.  \citet{Canuto91} suggest that a mixing length equal to the
distance to the boundary, $\alpha_{r-r_{\rm CZ}}=1$, should work for the entire
Sun. This does not match the actual mass mixing in the simulations
which is half of the expected value in the interior, adiabatic part of the
convection zone, and a factor of five larger near the surface.
From a relative logarithmic pressure of 0.5 and down to 4.7 where the bottom
boundary effects are felt, this normalization of the mass mixing length results
in a concave curve, and it is not obvious which range should be considered
close to constant. If we allow larger fluctuations around the mean, that mean
value will increase slightly as will the range over which the ratio is within
these fluctuations. To obtain a range similar to that of $\alpha_P$ would
require accepting fluctuations as large as $\pm 0.27$, apart from them not
being random fluctuations, but rather a systematic concaveness.

The reasoning behind choosing $\ell = r-r_{\rm CZ}$ is based on the MLT concept
of convective elements having size $\ell$, and that they cannot be larger than
their distance from the top of the convection zone.  Convection simulations show,
however, that the convective topology is one of upward and downward moving streams
with continual entrainment of upward moving fluid into the downflows.
Convection is a continuous not a discrete process.
There is therefore no conceptual problem with an $\ell > r-r_{\rm CZ}$.

As $\alpha_P$
is the most constant measure of the mass mixing length in the interior of the
deep solar simulation (Figure~\ref{fig:sun}), and we have similar experiences
with other simulations, we advocate a mixing length
formulation using $\alpha_P$. We note that with an adiabatic stratification in
hydrostatic equilibrium, $H_P = H_\rho/\gamma_1$. So $H_P < H_\rho$ due to the
simultaneous variation of $\rho$ and $T$, and consequently $\alpha_P \simeq
\gamma_1\alpha_\rho$. In the deep solar simulation $\gamma_1$ reaches the
completely ionized value of 5/3, but for the other 26 simulations, listed in
Table~\ref{tbl:mixinglength}, only the six simulations along the
high-$T_{\rm eff}$/low-$g$ edge get close to fully ionized, and the rest do
not get closer than $\gamma_1 \sim $1.22--1.32 (their minima are in the
1.16--1.22 range). The radial variation of $\gamma_1$ therefore also affects
the differences between mixing length normalizations with $H_\rho$ or $H_P$,
although the small gradient in the interior makes this a minor effect.

For other stars the available simulations are shallower (about 3\,dex less in
pressure), so some of the simulations have a smaller range
between the surface effects and the bottom boundary effects, and the
mixing length for those simulations are, for the
moment, less precisely determined.
 
\begin{figure}[!htb]
\centerline{\includegraphics[width=0.9\textwidth]{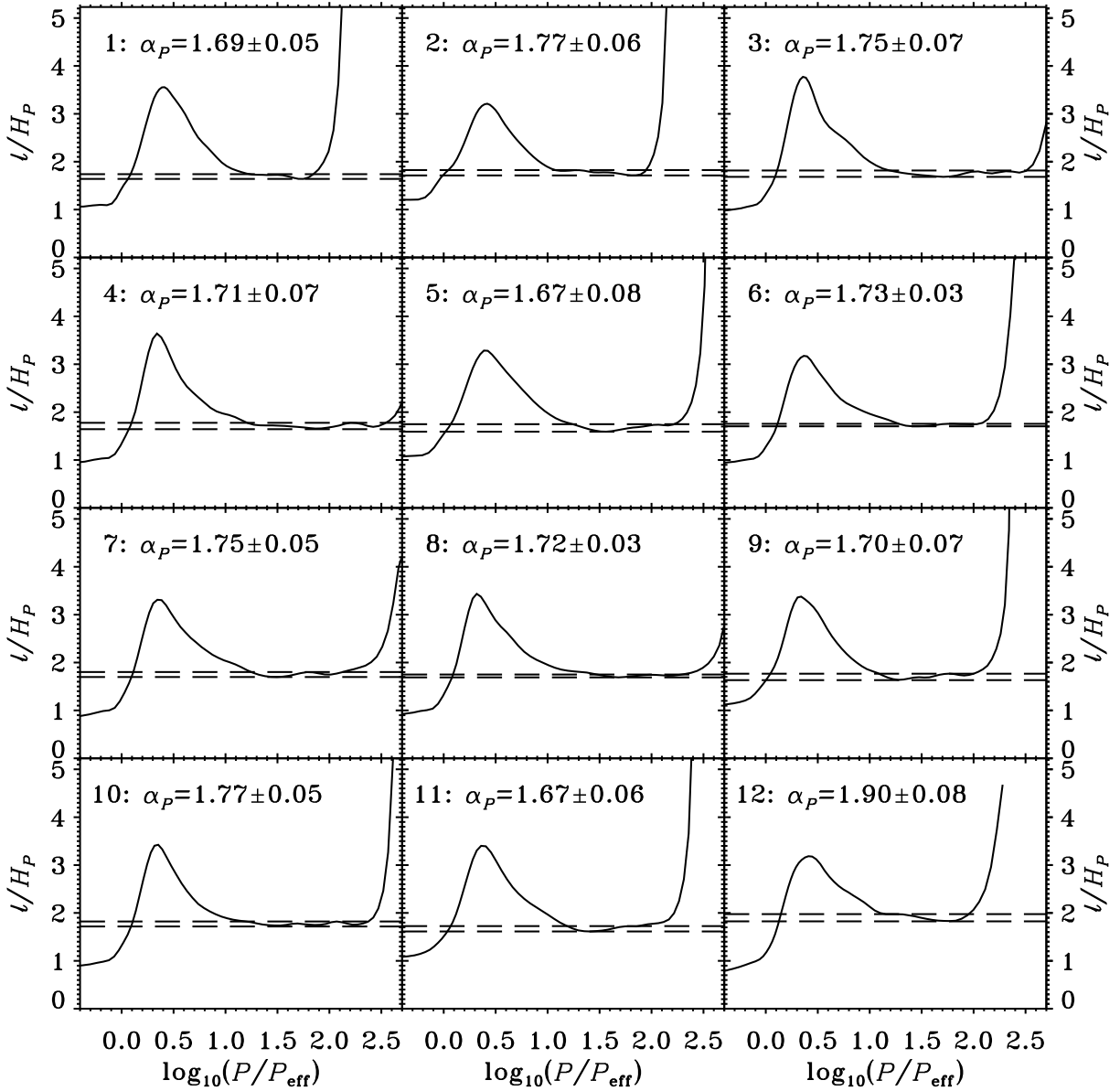}}
\caption{Ratio of mass mixing length to pressure scale height for stars
labeled by their simulation number in the table.  Solid line is the
calculated mass mixing length divided by the pressure scale height.
Dashed lines are the minimum and maximum range chosen for the ratio
$\alpha=\ell/H_P$.
} 
\label{fig:set1} 
\end{figure}

\begin{figure}[!htb]
\centerline{\includegraphics[width=0.9\textwidth]{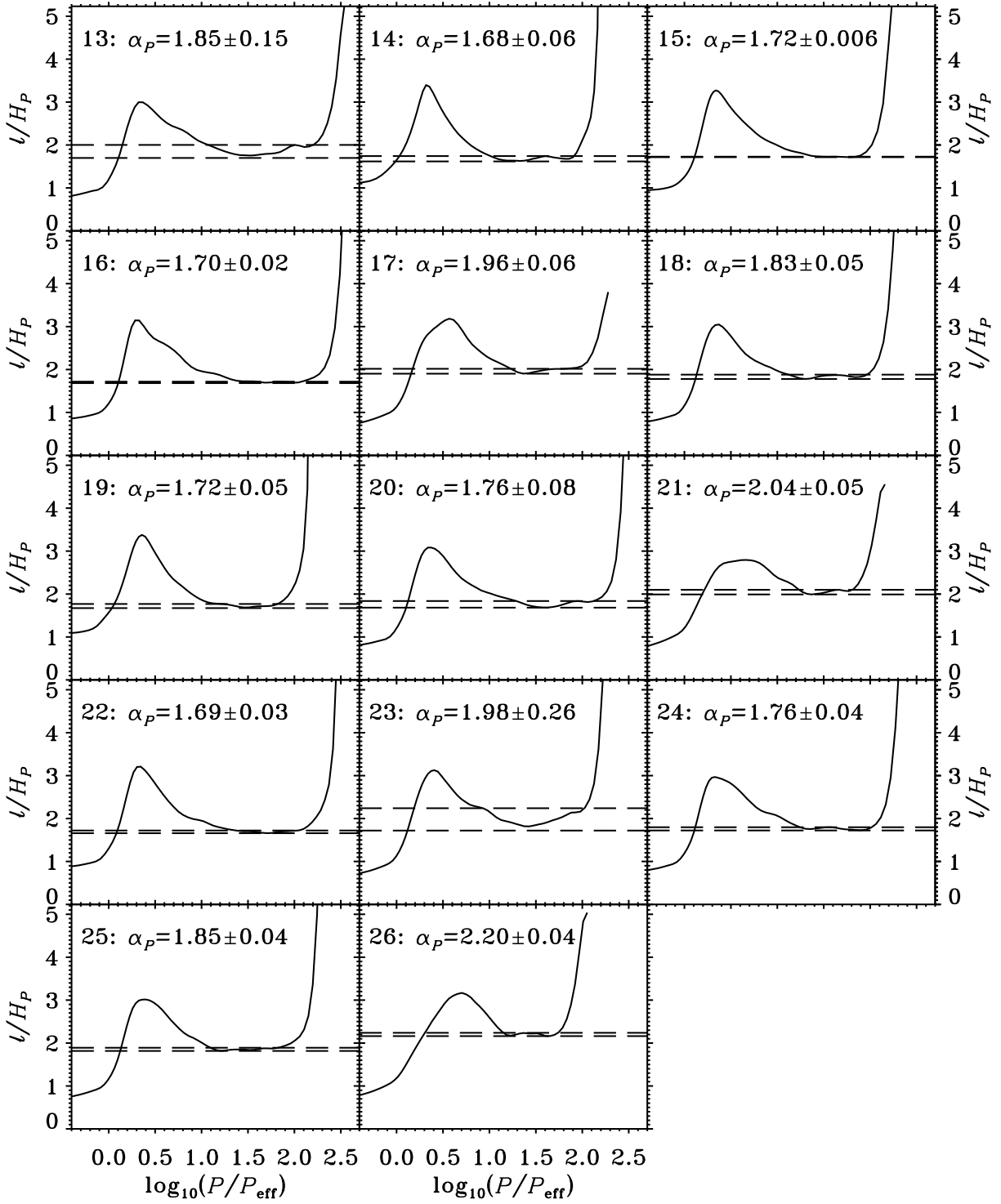}}
\caption{Next set of stars (See Fig.~\ref{fig:set1} for explanation).}
\label{fig:set2} 
\end{figure}

\begin{figure}[!htb]
\centerline{\includegraphics[width=0.9\textwidth]{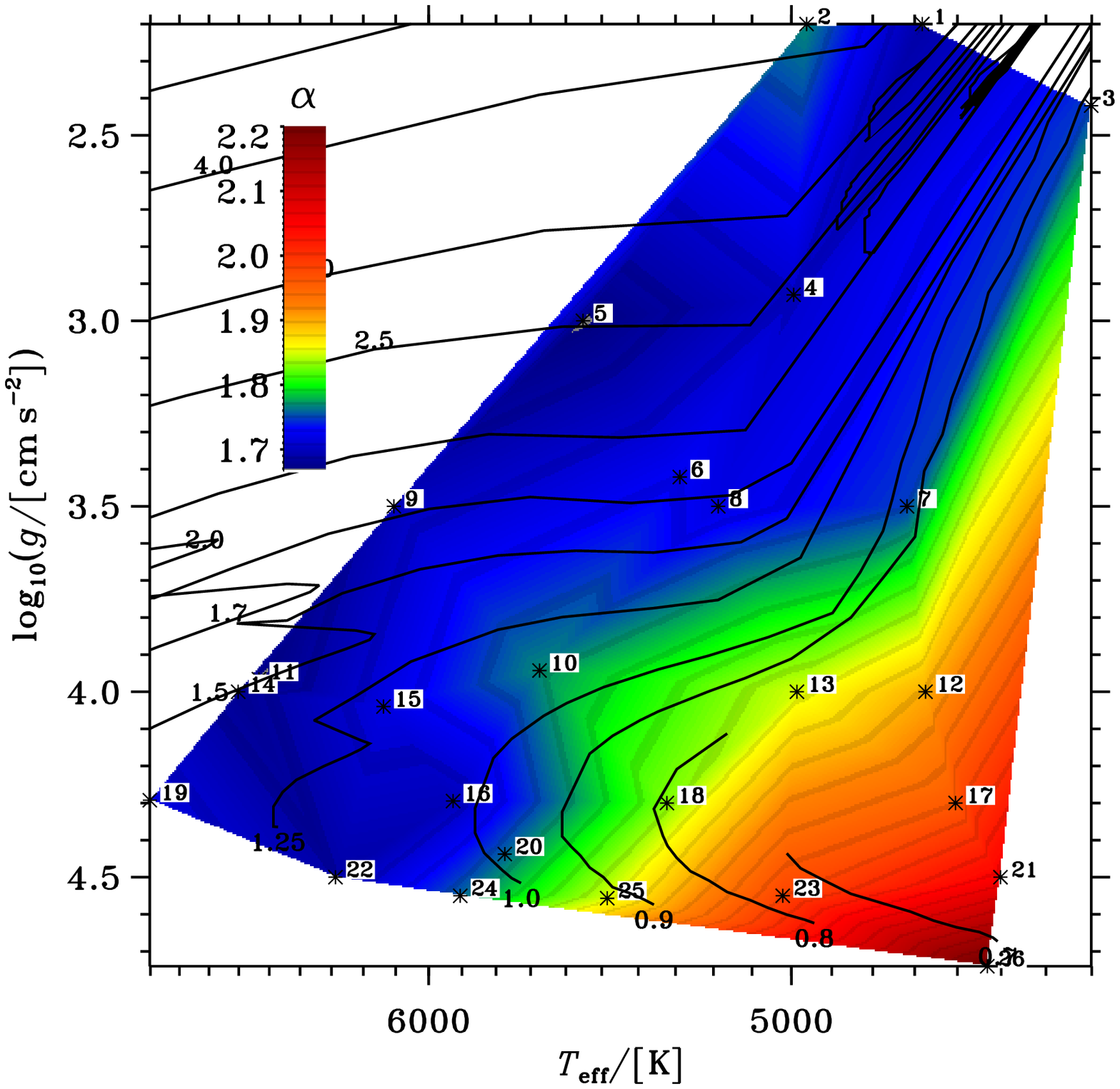}}
\caption{The mass mixing-length in units of $H_P$ for the 26 simulations, as
listed in Table~\ref{tbl:mixinglength} in the $T_{\rm eff}$ -- log g plane. We have also plotted evolutionary
tracks \citep{schaeller:stelev0.8-120,charbonnel:stelev0.4-1.0} for masses
as indicated at the bottom of each track.}
\label{fig:cmalfa} 
\end{figure}

Table~\ref{tbl:mixinglength} presents the mass mixing length in units of $H_P$
for the 26 stars for which we have performed convection simulations
and for which the simulations have a deep enough domain to determine
the mass mixing length. The table is ordered in order of increasing $\log g$,
and for similar gravity, in order of increasing $T_{\rm eff}$.
Figure~\ref{fig:set1}--\ref{fig:set2} show
the individual $\alpha_{\rm P}$ determinations and Figure~\ref{fig:cmalfa}
shows how $\alpha_{\rm P}$ varies as a function of effective temperature 
and gravity.

\begin{table}[!htb]
\caption{Mass mixing length and fundamental parameters for the 26 simulations}
{\begin{tabular}{rcccccc}
\hline
Simulation & MK\,class & \oh{$T_{\rm eff}$/[K]} & \oh{$\log g$} & \oh{$M/M_\odot$} & $\ell/H_P$ \\
\hline
  1 &     K3 &  4669 & 2.200 & 3.450 & $1.69\pm0.05$ \\
  2 &     K2 &  4962 & 2.200 & 4.300 & $1.77\pm0.06$ \\
  3 &     K5 &  4294 & 2.420 & 0.842 & $1.75\pm0.07$ \\
  4 &     K1 &  4994 & 2.930 & 2.350 & $1.71\pm0.07$ \\
  5 &     G8 &  5556 & 3.000 & 2.500 & $1.67\pm0.08$ \\
  6 &     K0 &  5288 & 3.421 & 1.833 & $1.73\pm0.03$ \\
  7 &     K3 &  4718 & 3.500 & 0.940 & $1.75\pm0.05$ \\
  8 &     K0 &  5189 & 3.500 & 1.660 & $1.72\pm0.03$ \\
  9 &     F9 &  6111 & 3.500 & 1.748 & $1.70\pm0.07$ \\
 10 &     G6 &  5671 & 3.943 & 1.110 & $1.77\pm0.05$ \\
 11 & Procyon F4 &  6529 & 3.966 & 1.500 & $1.67\pm0.06$ \\   
 12 &     K3 &  4673 & 4.000 & 0.727 & $1.90\pm0.08$ \\
 13 &     K2 &  4980 & 4.000 & 0.825 & $1.85\pm0.15$ \\
 14 &     F4 &  6611 & 4.000 & 1.490 & $1.68\pm0.06$ \\
 15 &     F9 &  6147 & 4.040 & 1.180 & $1.72\pm0.006$ \\
 16 &     G1 &  5929 & 4.295 & 1.028 & $1.70\pm0.02$ \\
 17 &     K4 &  4603 & 4.300 & 0.502 & $1.96\pm0.06$ \\
 18 &     K0 &  5327 & 4.300 & 0.788 & $1.83\pm0.05$ \\
 19 &     F2 &  6908 & 4.300 & 1.396 & $1.72\pm0.05$ \\
 20 & Sun G5 &  5778 & 4.438 & 1.000 & $1.76\pm0.08$ \\
 21 &     K4 &  4500 & 4.500 & 0.553 & $2.05\pm0.05$ \\
 22 &     F7 &  6286 & 4.500 & 1.201 & $1.69\pm0.03$ \\
 23 &     K1 &  5021 & 4.550 & 0.767 & $1.98\pm0.26$ \\
 24 &     G2 &  5894 & 4.550 & 1.085 & $1.76\pm0.04$ \\
 25 &     G9 &  5480 & 4.557 & 0.933 & $1.85\pm0.04$ \\
 26 &     K4 &  4531 & 4.740 & 0.800 & $2.20\pm0.04$ \\
\hline

\end{tabular}}
\label{tbl:mixinglength}
\end{table}


\section{Discussion and Summary}

The basic structure of stellar convection zones is determined by
the need to conserve mass.  Stars are highly stratified so most of
the mass ascending from depth must turn over and go back down
within the order of a scale height.  The high density mass from the
interior can not be supported in the low density outer layers.  This
produces a true mixing by entrainment of the almost laminar upflows into the
turbulent downflows.  There is therefore a real physical basis for the mixing
length theory of convection.  From computational fluid dynamic
simulations of stellar convection zones one can calculate this real
mixing of mass to determine the length scale on which it occurs.
Naively, one might expect it to be some multiple of the density scale
height.  However, as we observed for the Sun in Figure~\ref{fig:sun}, it is actually the
ratio of the mass mixing length to the pressure scale height which
is nearly constant over many orders of magnitude change in the
pressure. All of our values of mass mixing lengths are therefore presented
in terms of the pressure scale height, $H_P$.

In Figure~\ref{fig:cmalfa} we present the variation of $\alpha=\ell/H_P$
with the atmospheric parameters $T_{\rm eff}$ and $\log g$. First
off, comparing the range of $\alpha$ with the uncertainties in
Table~\ref{tbl:mixinglength}, we see that the variation over the HR diagram
is small, but significant.  The range of values found here (1.6-2.2) is 
consistent with those used in stellar evolution calculations 
\citep{Salaris08,Montalban06,Montalban04,Fernandes98,Chieffi95,Pedersen90},
although such calculations assume a constant value for $\alpha$ as the
star evolves.  We have also compared the mass mixing length values
found here with those determined by matching the adiabat in the
convection zone from the simulations with that obtained from mixing
length theory stellar models \citep{Trampedach11}. We find excellent agreement except
for the coolest, low mass, dwarf stars where the mass mixing length
values are larger.  It should also be remembered that in addition
to the other mixing length parameters specifying geometry and
radiative energy exchange, other input physics, atmospheric opacities
and the T-$\tau$ relation, also influence the appropriate value of
the mixing length to use in stellar structure and evolution
calculations.

We find $\alpha$ is smallest
towards the high-$T_{\rm eff}$ and low-$g$ side of the diagram.
These are also the simulations that have the least efficient
convection, where high speeds and large superadiabatic gradients
are necessary for carrying the flux to the photosphere. The least
efficient convection has the smallest mass mixing lengths,
as expected.  The most efficient mixing occurs in the coolest dwarf
with the most sedate convection. The solar value from our grid of
simulations of $\alpha_\odot = 1.76\pm 0.08$ agrees well with 
several determinations from the
literature, lending credence to the solar models and to the common
practice of calibrating $\alpha_{\rm MLT}$ against such models. We
also see from Figure~\ref{fig:cmalfa}, however, that the solar value
is less applicable for lower mass stars, which also have the largest
change of $\alpha$ during their evolution. The evolution of higher-mass
stars and the ascent up the Hayashi-track, seems well described by
a single value of $\alpha$.

The surface layers cannot be properly accounted for with a MLT-like
formulation. Crucial components like over-shooting, turbulent
pressure, radiative transfer in the presence of large temperature
inhomogeneities, etc., are not included in this simple model, but
greatly affect the surface layers.
The mass mixing length found here, is the appropriate mixing length
to use in a formulation that can realistically include all these phenomena
--- an ideal MLT formulation. Still lacking such a formulation, the solar mass
mixing length found in the present work, does not necessarily agree
with that found from matching solar evolution models to the Sun's
present luminosity and radius.  The latter furthermore depends on
all the other ingredients going into the atmosphere of the evolution
model, such as atmospheric opacities and how the photospheric
transition from optically thick to optically thin is treated, e.g.,
through some match to an atmosphere model or through $T$-$\tau$
relations. The mixing length of MLT is strongly connected to the
actual length-scale of mixing, as found in this paper, but this connection
is weakened by all these confounding factors entering into the MLT version.

The increase in the ratio of mass mixing length to pressure scale
height in the superadiabatic layers near the surface raises the
question of why stellar evolution calculations with the interior
values of $\alpha$ work well in matching observations of global stellar properties.  
In this regard, note that to match line profiles (e.g. Balmer lines
) a small
value of $\alpha$ that produces a steep temperature gradient is
needed in the atmosphere.  In this case, a large value of $\alpha$
must be used in the interior \citep{Montalban04}.  Our result suggest
such a two value set of mixing length parameters may indeed be
appropriate.  However, the main conclusion should be that mixing
length theory is not well defined and different variations are
possible that will yield similar results.

Calculations, such as presented here, will reduce the uncertainty
in the appropriate choice of mixing length in stellar structure and
evolution calculations.
In particular, we have used the behavior with
depth, to discriminate between various mixing length formulations, and
find that the mass mixing length, $\ell$, is best described with a constant
$\alpha_P = \ell/H_P$.
In a future paper \citep{Trampedach11} we present calibrations of
$\alpha_P({\rm MLT})$ against the simulations, based on 1D MLT envelope models
that employ the exact same EOS and opacities as the simulations, and use a new
self-consistent treatment of $T$-$\tau$ relations, computed from the
simulations \citep{trampedach:T-tau}. That calibration will be aimed at
reproducing the asymptotic adiabat with a standard MLT formulation, and will
contain our recommended values for $\alpha_P({\rm MLT})$.
Our present calculation of $\alpha$ is model independent in that we compute
the actual physical mixing length of the simulations, without connecting it
to a particular MLT formulation. Our present result, that $\alpha_P$ seems
convergent with depth, is a confirmation that MLT is a physically reasonable
model for stellar convection, despite its many shortcomings, particularly near
the surface. 
The present work also confirms the
soundness of the scheme employed in the \citet{Trampedach11}
$\alpha_P({\rm MLT})$ calibration, since, for most
of the simulations, it results in $\alpha$-values very similar to what we
find here.
This agreement between the two independent methods, suggests that
$\alpha_P({\rm MLT})$ has been approximately disentangled from
the various uncertainties in atmospheric physics that have historically
been absorbed into $\alpha_{\rm MLT}$.

With the results from \citet{trampedach:T-tau,Trampedach11}, supported by our present
results, 1D MLT models can be
made more self-consistent and with a reduced set of free parameters.
From an MLT point of view, there are of course still the undetermined geometric
form factors that could conceivably be adjusted to match the adiabat in
conjunction with the mass mixing length found here.

\acknowledgements

The calculations reported here were performed at the Australian Partnership
for Advanced Computations (APAC) and at the NASA Advanced Supercomputing (NAS)
Division of the Ames Research Center.  RT was supported by the Australian
Research Council (grants DP\,0342613 and DP\,0558836) and NASA grant NNX08AI57G.
RFS was supported by
NASA grants NNX07AO71G, NNX07AH79G and NNX08AH44G, and NSF grant AST0605738.
This support is greatly appreciated.


\begin{thebibliography}{27}
\expandafter\ifx\csname natexlab\endcsname\relax\def\natexlab#1{#1}\fi

\bibitem[{B{\"o}hm-Vitense(1958)}]{Vitense58}
B{\"o}hm-Vitense, E. 1958, \zap, 46, 108

\bibitem[{{Canuto} \& {Mazzitelli}(1991)}]{Canuto91}
{Canuto}, V.~M., \& {Mazzitelli}, I. 1991, \apj, 370, 295

\bibitem[{Charbonnel {et~al.}(1999)Charbonnel, D{\"a}ppen, Schaerer,
  Bernasconi, Maeder, Meynet, \& Mowlavi}]{charbonnel:stelev0.4-1.0}
Charbonnel, C., D{\"a}ppen, W., Schaerer, D., Bernasconi, P.~A., Maeder, A.,
  Meynet, G., \& Mowlavi, N. 1999, A\&APS, 135, 405

\bibitem[{{Chieffi} {et~al.}(1995){Chieffi}, {Straniero}, \&
  {Salaris}}]{Chieffi95}
{Chieffi}, A., {Straniero}, O., \& {Salaris}, M. 1995, \apjl, 445, L39

\bibitem[{{Fernandes} {et~al.}(1998){Fernandes}, {Lebreton}, {Baglin}, \&
  {Morel}}]{Fernandes98}
{Fernandes}, J., {Lebreton}, Y., {Baglin}, A., \& {Morel}, P. 1998, \aap, 338,
  455

\bibitem[{{Fernandes} {et~al.}(2002){Fernandes}, {Morel}, \&
  {Lebreton}}]{Fernandes02}
{Fernandes}, J., {Morel}, P., \& {Lebreton}, Y. 2002, \aap, 392, 529

\bibitem[{{Ferraro} {et~al.}(2006){Ferraro}, {Valenti}, {Straniero}, \&
  {Origlia}}]{Ferraro06}
{Ferraro}, F.~R., {Valenti}, E., {Straniero}, O., \& {Origlia}, L. 2006, \apj,
  642, 225

\bibitem[{{Freytag} \& {Salaris}(1999)}]{Freytag99}
{Freytag}, B., \& {Salaris}, M. 1999, \apjl, 513, L49

\bibitem[{Gough(1977)}]{gough:state-of-MLT}
Gough, D.~O. 1977, in {IAU Coll.} 38, Lecture Notes in Physics, Vol.~71,
  Problems of stellar convection, ed. E.~A. Spiegel \& J.~P. Zahn (Berlin:
  Springer), 15--56

\bibitem[{Gustafsson {et~al.}(1975)Gustafsson, Bell, Eriksson, \&
  Nordlund}]{gus:modgrid}
Gustafsson, B., Bell, R.~A., Eriksson, K., \& Nordlund, {\AA}. 1975, A\&A, 42,
  407

\bibitem[{Hummer \& Mihalas(1988)}]{mhd1}
Hummer, D.~G., \& Mihalas, D. 1988, ApJ, 331, 794

\bibitem[{{Iben}(1967)}]{Iben67}
{Iben}, Jr., I. 1967, \araa, 5, 571

\bibitem[{Kurucz(1992)}]{kur:line-data}
Kurucz, R.~L. 1992, Rev. Mex. Astron. Astrofis., 23, 45

\bibitem[{{Ludwig} {et~al.}(1999){Ludwig}, {Freytag}, \& {Steffen}}]{Ludwig99}
{Ludwig}, H., {Freytag}, B., \& {Steffen}, M. 1999, \aap, 346, 111

\bibitem[{{Montalb{\'a}n} \& {D'Antona}(2006)}]{Montalban06}
{Montalb{\'a}n}, J., \& {D'Antona}, F. 2006, \mnras, 370, 1823

\bibitem[{{Montalb{\'a}n} {et~al.}(2004){Montalb{\'a}n}, {D'Antona}, {Kupka},
  \& {Heiter}}]{Montalban04}
{Montalb{\'a}n}, J., {D'Antona}, F., {Kupka}, F., \& {Heiter}, U. 2004, \aap,
  416, 1081

\bibitem[{{Nordlund}(1976)}]{Nordlund76}
{Nordlund}, A. 1976, \aap, 50, 23

\bibitem[{{Pedersen} {et~al.}(1990){Pedersen}, {Vandenberg}, \&
  {Irwin}}]{Pedersen90}
{Pedersen}, B.~B., {Vandenberg}, D.~A., \& {Irwin}, A.~W. 1990, \apj, 352, 279

\bibitem[{{Salaris} \& {Cassisi}(2008)}]{Salaris08}
{Salaris}, M., \& {Cassisi}, S. 2008, \aap, 487, 1075

\bibitem[{Schaller {et~al.}(1992)Schaller, Schaerer, Meynet, \&
  Maeder}]{schaeller:stelev0.8-120}
Schaller, G., Schaerer, D., Meynet, G., \& Maeder, A. 1992, A\&APS, 96, 269

\bibitem[{{Stassun} {et~al.}(2004){Stassun}, {Mathieu}, {Vaz}, {Stroud}, \&
  {Vrba}}]{Stassun04}
{Stassun}, K.~G., {Mathieu}, R.~D., {Vaz}, L.~P.~R., {Stroud}, N., \& {Vrba},
  F.~J. 2004, \apjs, 151, 357

\bibitem[{{Stein} {et~al.}(2009){Stein}, {Georgobiani}, {Schafenberger},
  {Nordlund}, \& {Benson}}]{Stein09}
{Stein}, R.~F., {Georgobiani}, D., {Schafenberger}, W., {Nordlund}, {\AA}., \&
  {Benson}, D. 2009, in Amer. Inst. of Phys. Conf. Ser., Vol. 1094, , 764--767

\bibitem[{Stein \& Nordlund(1989)}]{bob:conv-topology}
Stein, R.~F., \& Nordlund, {\AA}. 1989, ApJ, 342, L95

\bibitem[{{Stothers} \& {Chin}(1997)}]{Stothers97}
{Stothers}, R.~B., \& {Chin}, C. 1997, \apjl, 478, L103

\bibitem[{Trampedach {et~al.}(2011{\natexlab{a}})Trampedach,
  {Christensen-Dalsgaard}, Nordlund, Asplund, \& Stein}]{trampedach:T-tau}
Trampedach, R., {Christensen-Dalsgaard}, J., Nordlund, {\AA}., Asplund, M., \&
  Stein, R.~F. 2011{\natexlab{a}}, A\&A, (submitted)

\bibitem[{Trampedach {et~al.}(2011{\natexlab{b}})Trampedach,
  {Christensen-Dalsgaard}, Nordlund, Asplund, \& Stein}]{Trampedach11}
---. 2011{\natexlab{b}}, (in preparation)

\bibitem[{{Trampedach} {et~al.}(1999){Trampedach}, {Stein},
  {Christensen-Dalsgaard}, \& {Nordlund}}]{Trampedach99}
{Trampedach}, R., {Stein}, R.~F., {Christensen-Dalsgaard}, J., \& {Nordlund},
  {\AA}. 1999, in Astron. Soc. Pac. Conf. Ser., Vol. 173, Stellar Structure:
  Theory and Test of Connective Energy Transport, ed. {A.~Gimenez,
  E.~F.~Guinan, \& B.~Montesinos}, 233--236

\end{thebibliography}

\end{document}